\documentclass{aa}  
\usepackage[colorlinks=true,citecolor=blue]{hyperref}
\usepackage{graphicx}
\usepackage{txfonts}

\begin{document} 

   \title{Using the Gerchberg–Saxton algorithm to reconstruct non-modulated pyramid wavefront sensor measurements}

   \author{V. Chambouleyron
            \inst{1}
            \and
            A. Sengupta
            \inst{1}
            \and
            M. Salama
            \inst{1}
            \and
            M. van Kooten
            \inst{2}
            \and
            B. L. Gerard
            \inst{3}
            \and
            S. Y. Haffert
            \inst{4}
            \and
            S. Cetre
            \inst{5,6}
            \and
            D. Dillon
            \inst{1}
            \and
            R. Kupke
            \inst{1}
            \and
            R. Jensen-Clem
            \inst{1}
            \and
            P. Hinz
            \inst{1}
            \and
            B. Macintosh
            \inst{1}
          }

   \institute{University of California Santa Cruz, 1156 High St, Santa Cruz, USA\\
              \email{vchambou@ucsc.edu}
        \and
        Herzberg Astronomy and Astrophysics 5071 West Saanich Road Victoria, British Columbia V9E 2E7
        \and
             Lawrence Livermore National Laboratory, 7000 East Ave., Livermore, CA 94550, USA
        \and
            University of Arizona, Steward Observatory, Tucson, Arizona, United States
        \and
             Durham University
        \and
             Wakea consulting
             }
 
  \abstract
   {Adaptive optics (AO) is a technique to improve the resolution of ground-based telescopes by correcting, in real-time, optical aberrations due to atmospheric turbulence and the telescope itself. With the rise of Giant Segmented Mirror Telescopes (GSMT), AO is needed more than ever to reach the full potential of these future observatories. One of the main performance drivers of an AO system is the wavefront sensing operation, consisting of measuring the shape of the above mentioned optical aberrations.}
   {The non-modulated pyramid wavefront sensor (nPWFS) is a wavefront sensor with high sensitivity, allowing the limits of AO systems to be pushed. The high sensitivity comes at the expense of its dynamic range, which makes it a highly non-linear sensor. We propose here a novel way to invert nPWFS signals by using the principle of reciprocity of light propagation and the Gerchberg-Saxton (GS) algorithm.}
   {We test the performance of this reconstructor in two steps: the technique is first implemented in simulations, where some of its basic properties are studied. Then, the GS reconstructor is tested on the Santa Cruz Extreme Adaptive optics Laboratory (SEAL) testbed located at the University of California Santa Cruz.}
   {This new way to invert the nPWFS measurements allows us to drastically increase the dynamic range of the reconstruction for the nPWFS, pushing the dynamics close to a modulated PWFS. The reconstructor is an iterative algorithm requiring heavy computational burden, which could be an issue for real-time purposes in its current implementation. However, this new reconstructor could still be helpful in the case of many wavefront control operations. This reconstruction technique has also been successfully tested on the Santa Cruz Extreme AO Laboratory (SEAL) bench where it is now used as the standard way to invert nPWFS signal.}
   {}


\maketitle


\section{Introduction}

The pyramid wavefront sensor (PWFS) \citep{raga} falls under the category of Fourier-filtering wavefront sensors \citep{fauva2017}, which are commonly used to measure aberrations in optical systems. Inspired by the Foucault knife test, the original PWFS consists of a 4-sided glass pyramid located at an intermediate focal plane and a detector that captures images of the four pupils created by each beam passing through the different faces of the pyramid. This configuration efficiently converts phase into intensity, but lacks the necessary dynamic range to accurately measure atmospheric turbulence aberrations, which can induce optical path differences on the order of several waves. To address this issue, the PWFS is often paired with a modulator, which causes the electromagnetic (EM) field to circulate around the pyramid tip during the camera's acquisition time. Modulation drastically improves the PWFS's linearity, but it comes with three main drawbacks. First, the modulation leads to a strong decrease in PWFS sensitivity, especially for low-order modes. Secondly, modulating the PWFS alters the nature of the signal \citep{2004OptCo.233...27V,Guyon_2005}, causing the response to resemble that of a slope-sensor, hence making it difficult to detect phase discontinuities. This is particularly problematic for the next generation of Giant Segmented Mirror Telescopes (GSMTs), where wavefront control will involve correcting not only turbulence-induced aberrations but also those induced by the telescope itself (fragmentation \citep{schwartzSensingControlSegmented2017,ariellePetal,GMTphasing} and segmentation \citep{Chanan:98}). Finally, adding the modulation mirror stage requires the use of more optics and leads to difficulties related to fast steering components (stability issues, temperature constrains, speed limitations, failure risks, etc...). Therefore, extending the dynamic range of the non-modulated PWFS (nPWFS) to remove the need for modulation could allow the use of the PWFS at its full potential while removing the requirement for moving parts, making it a less complex system. Already a lot of studies have been done to deal with PWFS non linearities and several options envisaged. It was proposed to keep the matrix formalism and consider the PWFS as a linear varying -parameter system: the reconstruction matrix evolves according to the phase to be measured \cite{Korkiakoski2008,Deo2019}, but this technique usually requires some knowledge on the statistics of the measured phase \citep{Chambouleyron2020}. Gradient-descent methods have also been investigated \citep{Frazin:18,Hutterer_2023}. Finally, another approach, more appealing today, is to use a machine-learning approach to reconstruct the nPWFS signal \citep{Landman:20,2022A&A...664A..71N,2023ML}.

The goal of this paper is to present a novel approach to invert the nPWFS signal. This approach is based on the principle of reciprocity of light propagation and the Gerchberg-Saxton (GS) algorithm. In section 2, we will present in detail the principle of the reconstruction algorithm. Section 3 will assess the basic performance of the reconstructor, mainly in terms of dynamic range but also in terms of noise propagation and broadband performance. Section 4 will highlight the results of an experimental implementation of this new reconstructor on the Santa Cruz Extreme Adaptive optics Laboratory (SEAL) testbed \citep{SEAL}. Finally, section 5 will describe a possible way to push even further the dynamic range and convergence speed of this reconstructor by using sensor-fusion.

\section{A new method to invert non-modulated PWFS signal}

\subsection{Reciprocity of light propagation principle}

The reconstructor presented in this paper relies on one of the most basic properties of light propagation: the principle of reciprocity of light propagation. The idea is to construct a high-fidelity numerical model of the nPWFS and use the measurements to send the light backwards in the numerical nPWFS. This technique could actually apply to any Fourier-filtering wavefront sensor, but we will focus only on the nPWFS throughout this study. 

The numerical model of the nPWFS is built by propagating the light assuming Fraunhofer approximation from one pupil plane to another, while going through the pyramid mask in an intermediate focal plane. In more details, the EM-field in the nPWFS detector plane can be written as:

\begin{equation}
    \Psi_{d} = \Psi_{p} \star \widehat{m}
    \label{eq:propag}
\end{equation}

where $\Psi_{p}$ is the EM-field in the entrance pupil, $m$ the complex shape of pyramid mask, $\star$ is the convolution product and $\widehat{\cdot}$ denotes the Fourier transform operator. Equation  \ref{eq:propag} allows us to simply simulate light propagation from the entrance pupil plane to the WFS detector plane. The back propagation of light from detector plane to the entrance pupil plane is:

\begin{equation}
    \Psi_{p} = \Psi_{d} \star \widehat{\bar{m}}
    \label{eq:Back_propag}
\end{equation}

where $\bar{\cdot}$ is the conjugate operator. This last equation can be simply understood in the nPWFS case: we write $m$ as $m=e^{i\Delta}$ where $\Delta$ is a 2D real function describing the phase corresponding to the pyramid shape. Going through the pyramid in the opposite direction means that the light propagates through the inverse phase mask, \textit{i.e} $\bar{m} = e^{-i\Delta}$. We can then write the entrance pupil and detector EM-fields in their complex form:

\begin{equation}
    \begin{cases}
        \Psi_{p} &= A_{p}e^{i\phi_{p}} \\
        \Psi_{d} &= A_{d}e^{i\phi_{d}} \\
        \end{cases}
        \label{eq:complexEM}
\end{equation}

where $A_{p}$ and $A_{d}$ are the amplitudes and $\phi_{p}$ and $\phi_{d}$ are the phases of the electromagnetic fields. The goal of wavefront sensing is to find back the phase $\phi_{p}$ in the entrance pupil. Light back propagation cannot be easily performed from the detector measurements because we only have access to intensities in the detector plane $I_{d}=|\Psi_{d}|^{2} = A_{d}^{2}$. The phase $\phi_{d}$ is therefore missing in the measurements. Hence, we propose to use an iterative algorithm called the Gerchberg-Saxton (GS) algorithm to propagate the light back and forth in the numerical model of the nPWFS.

\subsection{Gerchberg-Saxton algorithm}

\begin{figure*}
  \includegraphics[width=2\columnwidth]{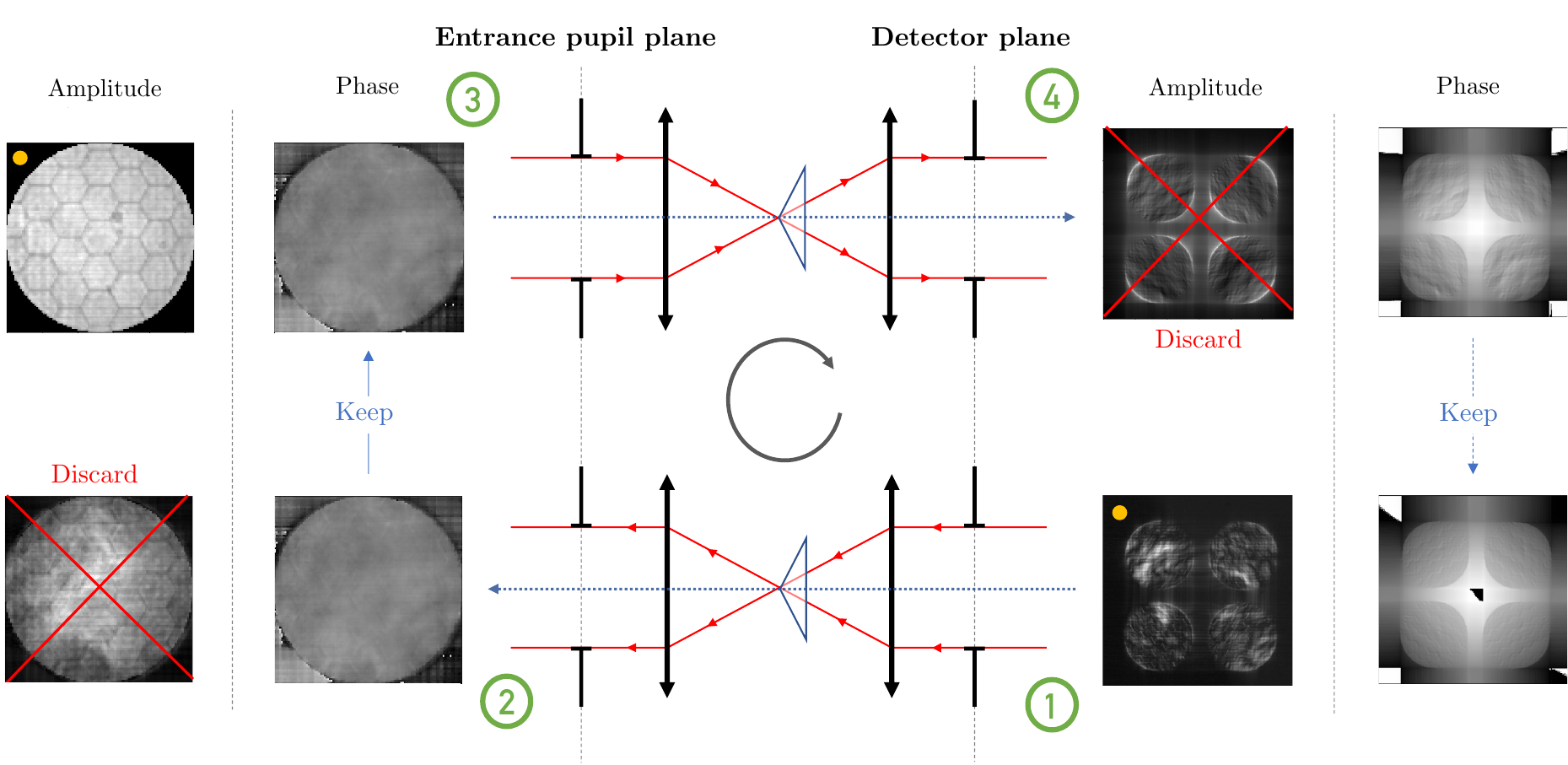}
  \caption{Principle of the GS algorithm applied to non-modulated PWFS. Yellow dots are highlighting the fact that these are true data obtained on the SEAL testbed.}
  \label{fig:GS_algo}
\end{figure*}

The GS algorithm was first proposed by \citep{gerchberg1972practical} and is widely used to perform image sharpening from point spread function (PSF) images \citep{Fienup:82,keck_GS}. To perform this algorithm in our case, we will assume that we have access to a measurement of the entrance pupil amplitude $A_{p}$ (we will show later an easy practical way to obtain this quantity with the nPWFS). In the nPWFS framework, we have two complex quantities (equation \ref{eq:complexEM}) for which the amplitudes are known and with a relation that links them together (equation \ref{eq:propag}). The principle of the GS algorithm is to propagate the light back and forth in the numerical model of the nPWFS, injecting at each step the knowledge of the amplitudes of the complex quantities we are trying to retrieve. We propose to go through one iteration of the GS algorithm applied to the nPWFS based on the schematic given figure \ref{fig:GS_algo}. The amplitude for the entrance pupil and the detector plane used for this example are true measurements from the SEAL testbed (highlighted with yellow dots). One iteration of the GS algorithm can be split in four parts:
\begin{enumerate}
    \item We compute the detector EM-field complex amplitude $A_{p} = \sqrt{I_{p}}$. For the first iteration, the complex EM-field in the detector plane is built by using $\phi_{d}=\arg(A_{p} \star \widehat{m})$, which corresponds to the phase at the detector plane when a flat wavefront is propagated through the nPWFS system. We therefore have a first estimation $\Psi_{d}$ that can be back-propagated in the system (through equation \ref{eq:Back_propag}).
    \item A first estimation of $\Psi_{p}$ is then obtained. 
    \item Since we already have access to $A_{p}$ the amplitude found through back-propagation is discarded and replaced by the measurement of $A_{p}$ while keeping the estimated phase $\phi_{p}$. The entrance pupil plane EM-field can then be propagated in the system (direct propagation, through equation \ref{eq:propag}).
    \item A new estimation of $\Psi_{d}$ is obtained. As previously done for the entrance pupil plane, we discard the amplitude and replace it by the measurement of $A_{d}$ given by the detector and keep the estimated phase $\phi_{d}$. We can then go back to step 1 and iterate again. In this paper, we will call one iteration of GS algorithm the numerical operation consisting in going through these four steps. 
\end{enumerate}

The GS algorithm is therefore an iterative algorithm, which for one iteration performs four Fast-Fourier Transforms. It is important to notice that this reconstructor assumes coherence of the EM-field. Therefore, this algorithm does not work for the modulated PWFS. It also raises the question of the impact of measurements with larger spectral bandpasses, which will be tackled in the next section. We also emphasize that this GS principle can be applied to any FFWFS, but in this paper we focus only on the nPWFS.

\subsection{Phase unwrapping}

The phase, $\phi_{p}$, retrieved by the algorithm will be wrapped, modulo $2\pi$, as optical propagation is done. Therefore when working with phase with amplitude larger than $2\pi$ peak-to-valley (PtV), it is necessary to add an extra step to the reconstruction process: a phase-unwrapping algorithm applied to the phase estimated through the GS reconstruction. A detailed analysis of this step is out of the scope of this study, hence we will restrain to using a well-known "off-the-shelf" algorithm based on \citep{Ghiglia:94} throughout all this paper.

\section{Performance of the GS algorithm reconstructor}
\label{section:perfo}
This section is dedicated to giving a first overview of the GS algorithm's basic performance. We will focus on how this reconstructor performs in terms of dynamic range, while also evaluating the number of iterations needed, noise propagation and broadband impact on reconstruction. This study is not meant to be exhaustive, and parameters like the minimum number of pixels needed on the detector with respect to the amplitude of the phase to be measured will not be assessed nor a fine analysis of the impact of model errors. All the simulations will simply match the SEAL testbed configuration. Here are the main simulation parameters:
\begin{itemize}
    \item[$\bullet$] Each pupil on the PWFS detector has 106 pixels across (matching SEAL testbed configuration), which is a realistic case as the best low read-out-noise (RON) cameras can run at a few kHz with resolutions around $250px\times250px$. We use a Shannon sampling of 2 (4 pixels per $\lambda/D$) for the nPWFS model.
    \item[$\bullet$] For a more realistic simulation, the phase screens are first simulated and propagated with a resolution 4 times bigger (4x106px across) through a high-resolution model of the nPWFS with also a Shannon sampling of 2. The signal is then binned to produce the nPWFS image with 106px across.
    \item[$\bullet$] We work with a modal basis of the first 500 Zernike modes (excluding piston), close to the control space that can be achieved with the deformable mirrors installed on the SEAL testbed.
    \item[$\bullet$] The linear reconstructor is created in a standard fashion by building an interaction matrix and computing its pseudo-inverse. No modes are filtered during the inversion as the measurements are over-sampled, leading to a good conditioning number of the interaction matrix.
    \item[$\bullet$] Simulations tools used are internally sourced, except for the turbulent phase screens which are created using the HCIpy library \citep{HCIpy}.
\end{itemize}

\subsection{Convergence speed and first comparison with linear reconstructor}

The first test presented aims at giving an idea of the convergence speed of the GS reconstruction. It will also be a way to draw a first comparison between the linear model and our reconstructor. For that, we are studying the reconstruction of four different turbulent phase screens following a Von-Karman spectrum and corresponding to four different configuration of $D/r_{0}$ where $D$ is the telescope diameter and $r_{0}$ the Fried parameter. To produce a fair comparison between the linear reconstructor and the GS reconstructor which gives the phase with a much higher resolution (106px across), these screens are projected on the 500 Zernike modes basis before propagation (no aliasing). The reconstruction error is estimated at each GS iteration (estimated phase is systematically unwrapped) and is expressed as a ratio of the rms error in the entrance pupil plane (error = $rms_{rec}/rms_{input}$).

\begin{figure}[h!]
\centering
	\includegraphics[width=1\columnwidth]{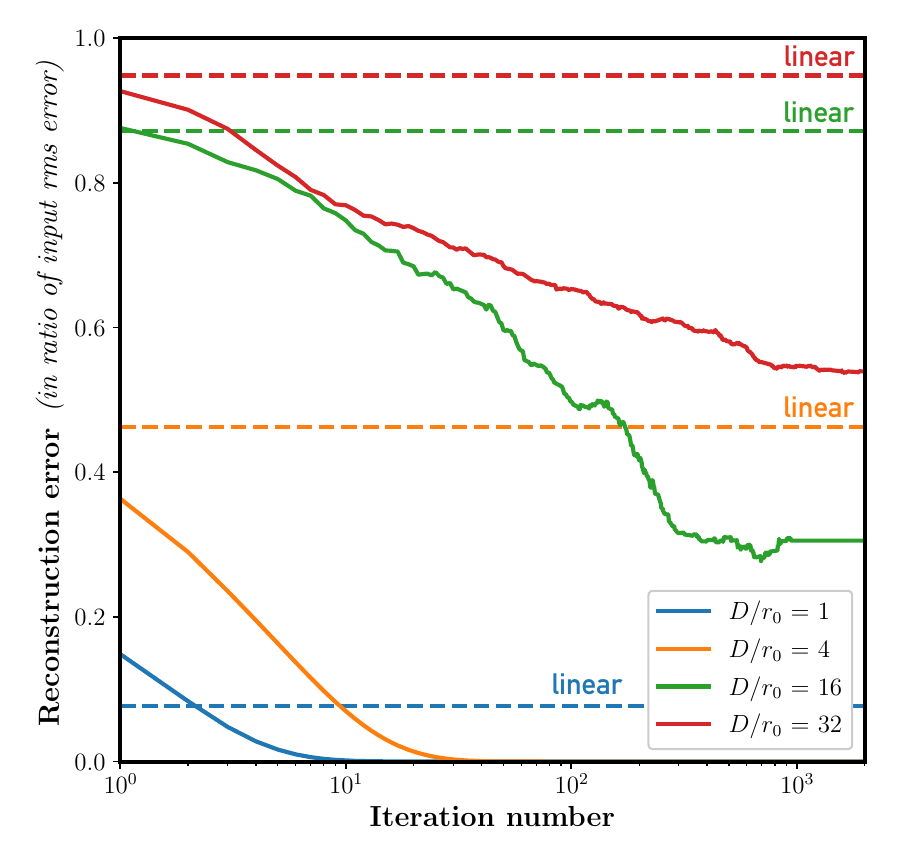}
    \caption{GS algorithm convergence speed. The GS algorithm is applied for input turbulent screens for four different seeing conditions. Dashed lines represent the corresponding reconstruction error in the linear framework.}
    \label{fig:convergence}
\end{figure}

\begin{figure}[h!]
\centering
	\includegraphics[width=1\columnwidth]{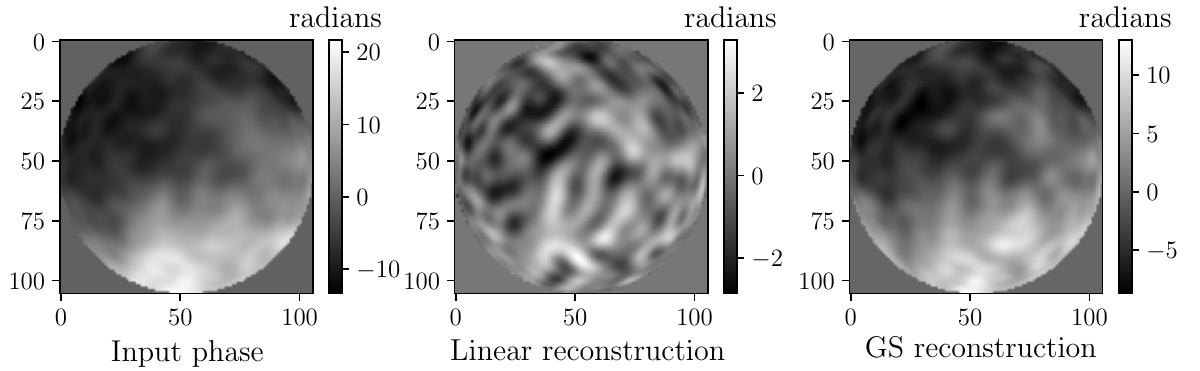}
    \caption{\textbf{Left:} Input turbulent phase in the case $D/r_{0} =32$, projected on the first 500 Zernike modes. \textbf{Middle:} Linear reconstruction. \textbf{Right:} GS reconstruction after 2000 iterations.}
    \label{fig:convergence_comparison}
\end{figure}

Results are given in figure \ref{fig:convergence}, where the horizontal dashed lines correspond to the linear reconstructor and the x-axis is given in log-scale. One can notice that the GS reconstruction accuracy increases with the number of iterations, before stabilizing. The smaller the phase, the less iterations are needed to reach the best reconstruction. Hence we see that for $D/r_{0} = 1$ only 10 iterations are required, whereas it needs more than 1000 iterations for the case $D/r_{0} = 32$ (corresponding to a value of $r_{0} = 25\ cm$ for a $D=8\ m$ telescope). This figure also shows that in any case and after a few iterations, the GS reconstructor performs better than the linear reconstructor. In the small phase regime (in our context, it would correspond to regimes smaller than $D/r_{0}\approx0.5$), both will perform evenly as the nPWFS would be working in its linear range. To illustrate reconstruction products, input phase and reconstructed phases (linear and GS after 2000 iterations) for the case $D/r_{0} = 32$ are shown in figure \ref{fig:convergence_comparison}. It is clear that even if the GS reconstruction still underestimates the phase (due to nPWFS saturation, this will be discussed in more details in section \ref{section:seal}), the shape of the retrieved phase is much closer than the one produced by the linear reconstructor. As a side note on the shape given by the linear reconstructor: the phase is highly underestimated and the pattern seems quite different than the input phase with larger high spatial frequencies, suggesting that it would be hard to start a closed loop with such a reconstruction. As the nPWFS is sensitive to phase discontinuities, the linear reconstruction shape could be partially explained by the impact of phase wrapping on measurements. A detailed analysis of this phase wrapping effect on the linear reconstruction is out of the scope of this paper, but remains an important point to understand nPWFS behavior.

\begin{figure}[h!]
\centering
	\includegraphics[width=1\columnwidth]{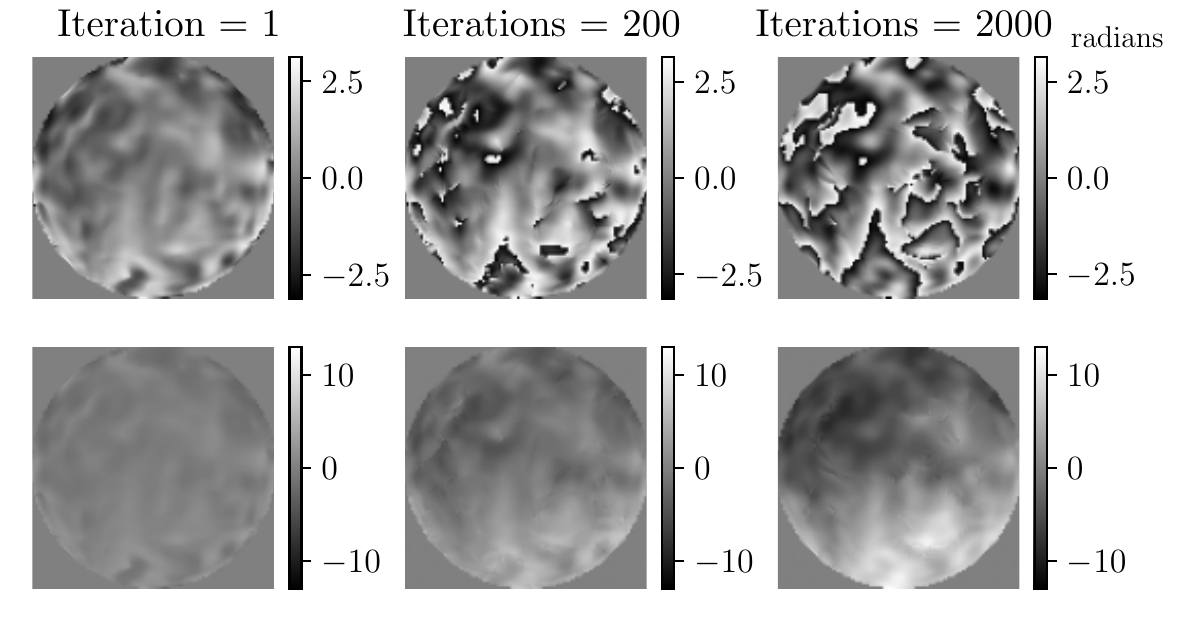}
    \caption{\textbf{Top:} Wrapped phases reconstructed by the GS algorithm. \textbf{Bottom:} Corresponding unwrapped phases.}
    \label{fig:convergence_phases}
\end{figure}

To illustrate how the GS reconstructed phases evolve with iteration, reconstructed phases for iterations 1, 200, and 2000 are shown in figure \ref{fig:convergence_phases} for the  strongest turbulent case $D/r_{0} = 32$. The top row is presenting the wrapped phase given by the GS algorithm and the bottom row shows the corresponding unwrapped phases. The algorithm seems to quickly converge towards the good overall wrapped shape and then improves the estimation by scaling the phase closer from the input amplitude (by therefore increasing the phase wrapping). For figure \ref{fig:convergence}, seeing limited phase screens were used as examples to give a first insight on GS reconstructor behavior. In a true AO system, the nPWFS would typically work in closed loop around residual phases, so typical AO residual phases could have also been used for this previous analysis. We instead decided to use a full power law turbulence phase screen because: (i) The closed loop bootstrapping is done on full turbulence anyways and (ii) using AO residual phase screens would have been required to choose a specific system configuration. In order to refine the comparison between linear and GS reconstructors, it was decided to build linearity curves for Zernike modes. Results are presented in the next section.

Finally, from this first analysis, it seems that the GS reconstructor outperforms the linear reconstructor, but it requires tens of iterations to be effective. Each propagation back and forth requires four 2D Fast-Fourier-Transforms (FFTs) (figure \ref{fig:GS_algo}). In the current implementation, \textit{numpy.fft.fft2} is used to compute the Fourier transforms on a problem size of [424, 424] (106 pixels across pupil, with a sampling of 2 times Shannon), averaging at $7.1 \pm 0.4$ ms for each FFT on the SEAL control computer. Although the 2D FFTs algorithm scales as $O(N^2 \log N)$ overall, problem sizes that are a factor of 2 or that are divisible by large powers of 2 or 3 can run significantly faster. Padding to $448 = 2^6 \times 7$ improves the \textit{numpy.fft.fft2} runtime to $3.2 \pm 0.07$ ms. With the downsampled problem size of [212, 212] (that could easily be reached by decreasing sampling and reducing the number of subapertures) and a more efficient FFT algorithm called \textit{FFTW} \citep{FFTW}, the runtime goes down to $640 \pm 27 \mu$s, an order of magnitude compared with the initial runtime. This represents an efficiency of about 5500 mflops, below the stated achievable performance for similar CPUs of $\sim 12000$ mflops. It is likely that an improved CPU would be able to achieve this performance, enabling us to run each FFT at the full problem size in less than 1 ms. Further, it is possible that a lower-level implementation of the GS algorithm would be able to make use of the repeated FFT per iteration, for example by optimizing memory access or by varying the FFT problem size per iteration. Based on prior benchmarking led for the CUDA-based \textit{cuFFT} implementation \citep{cuFFT}, GPU computation could reduce this up to an order of magnitude further. This could allow us to run few iterations of the algorithm within a millisecond, opening the path for closed-loop operation.

\subsection{Linearity Curves and dynamic range plots}

To achieve a more quantitative comparison in terms of dynamic range between the linear and the GS reconstructors, linearity curves for some Zernike modes are analyzed. Zernike modes within a full range of amplitudes are sent through the PWFS and then reconstructed, for four different cases: \textit{(i)} nPWFS with linear reconstructor \textit{(ii)} modulated PWFS with a modulation radius of $3\lambda/D$ as a comparison point in terms of dynamic range \textit{(iii)} nPWFS with GS reconstructor without the unwrapping step \textit{(iv)} nPWFS with GS reconstructor with phase unwrapping. 

\begin{figure}[h!]
\centering
\includegraphics[width=0.49\textwidth]{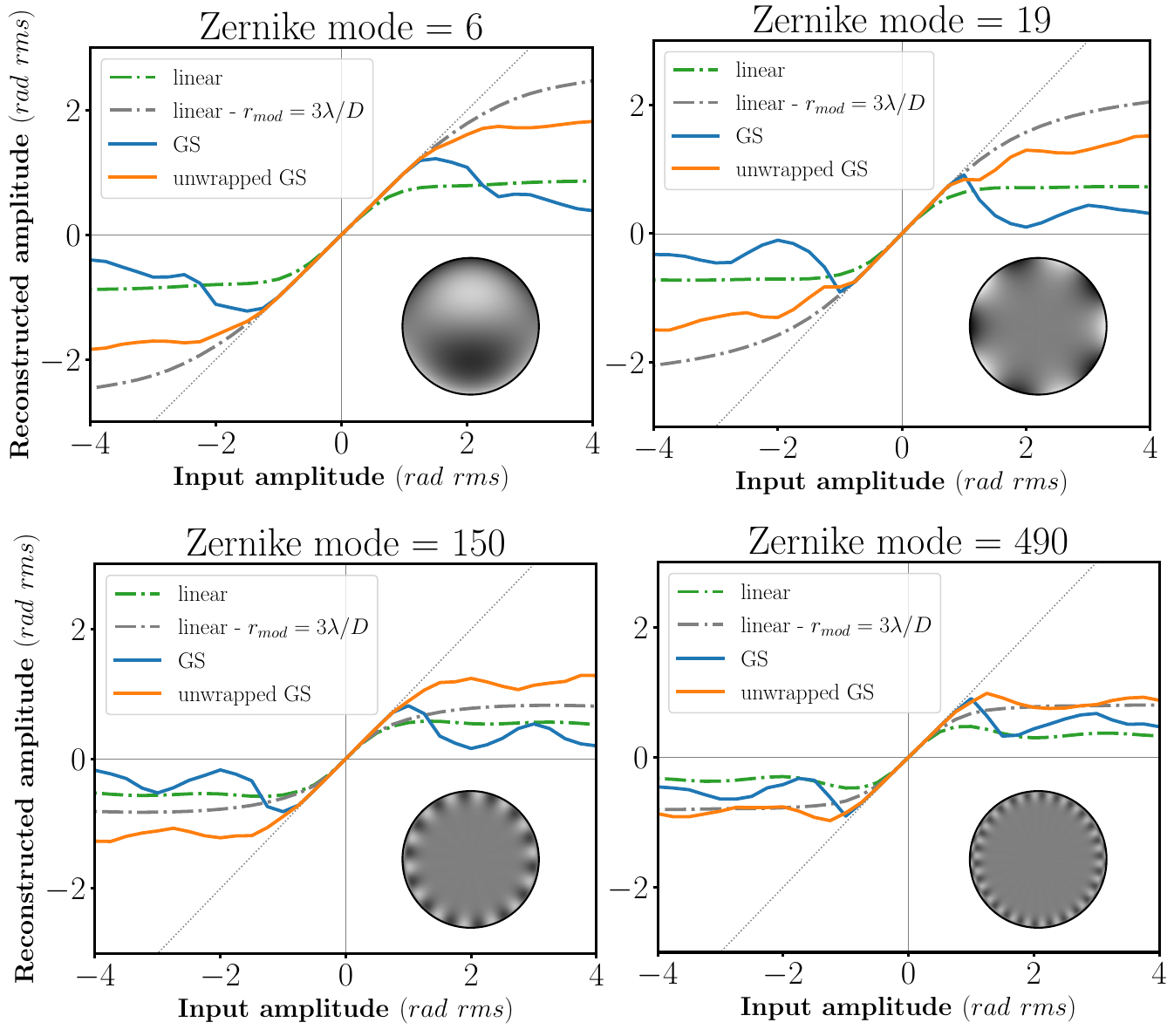}
\caption{Linearity curves for different Zernike modes, ranging from low order to high order. GS algorithm is run with 25 iterations in this case.}
\label{fig:linearityCurves}
\end{figure}

Results for four different Zernike modes ranging from low spatial frequency to high spatial frequency ($Z^{6}$, $Z^{19}$, $Z^{150}$ and $Z^{490}$) are given in figure \ref{fig:linearityCurves}. In this case, the GS reconstructor was used with an arbitrary number of 25 iterations (impact of number of iterations will be assessed further in this section). These linearity curves are showing the well known behavior of the modulated PWFS with respect to the nPWFS: significant increase in dynamic range for low order modes located within the modulation radius, and comparable dynamic range for high-order modes composed of spatial frequencies outside the modulation radius. The GS algorithm without the unwrapping algorithm is showing extended linear range for all the modes, but the response curves are steeply dropping after 1 rad rms of aberrations. It actually corresponds to the amplitude for which the phase starts to wrap (PtV greater than $2\pi$), hence the reconstructed phase starts to have a different shape from the unwrapped phase and projection on Zernike modes is modified as more high-order modes are introduced. However, adding the unwrapping step to the reconstruction allows us to better improve the linearity curves after 1 rad rms of input phase. It is clearly demonstrated here that the GS reconstructor added with the unwrapping step allows us to significantly increase the nPWFS dynamic range for all modes.

\begin{figure}[h!]
\centering
\includegraphics[width=0.5\textwidth]{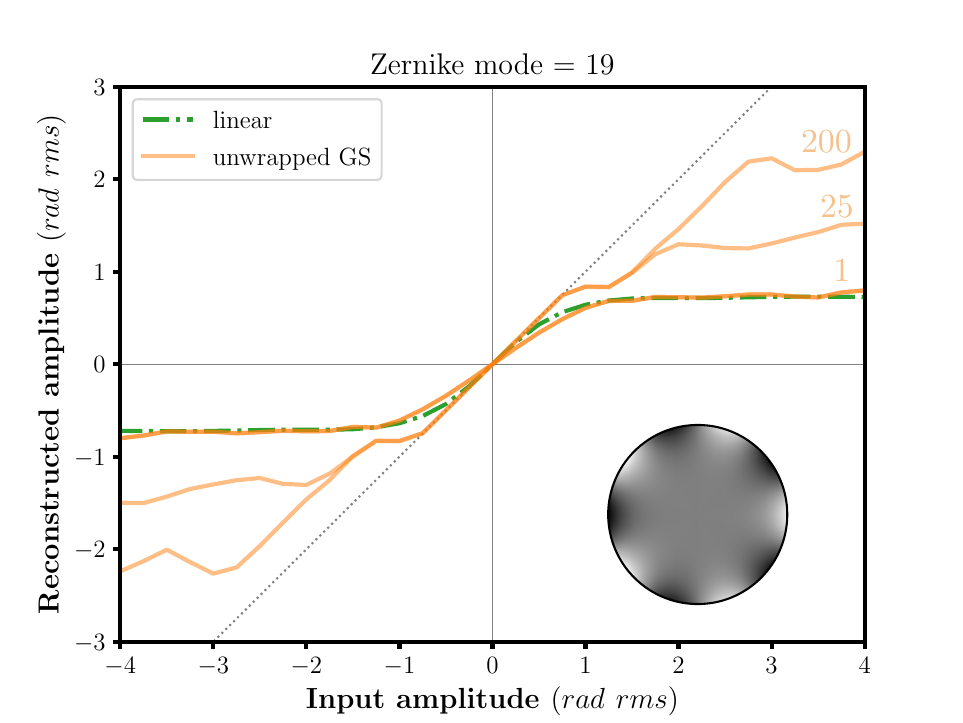}
\caption{Linearity curves for the Zernike mode 19 for different numbers of iterations while performing the GS algorithm. Number of iterations are written on the top-right part of the curves.}
\label{fig:linearityCurves_iterations}
\end{figure}

For the plots in figure \ref{fig:linearityCurves}, it was arbitrarily chosen to run 25 iterations for the GS algorithm. The impact of the number of iterations on the linearity curve of the Zernike mode $Z^{19}$ is assessed in figure \ref{fig:linearityCurves_iterations} where the linearity curve for this mode is shown using 1, 25 and 200 iterations in the case of GS algorithm combined with the unwrapping step. One can notice that for 1 iteration, the linearity curve doesn't follow the $y=x$ slope around a null phase. That shows that even in a small phase regime, only one iteration of the algorithm is not enough to accurately find back the phase. We noticed that actually only 2 iterations are enough in the small phase regime (in the ideal case of these simulations where there is no model error). Then, as expected the linearity curves improve with the number of iterations, especially for the strong aberrations regime.

\begin{figure}[h!]
\centering
\includegraphics[width=0.5\textwidth]{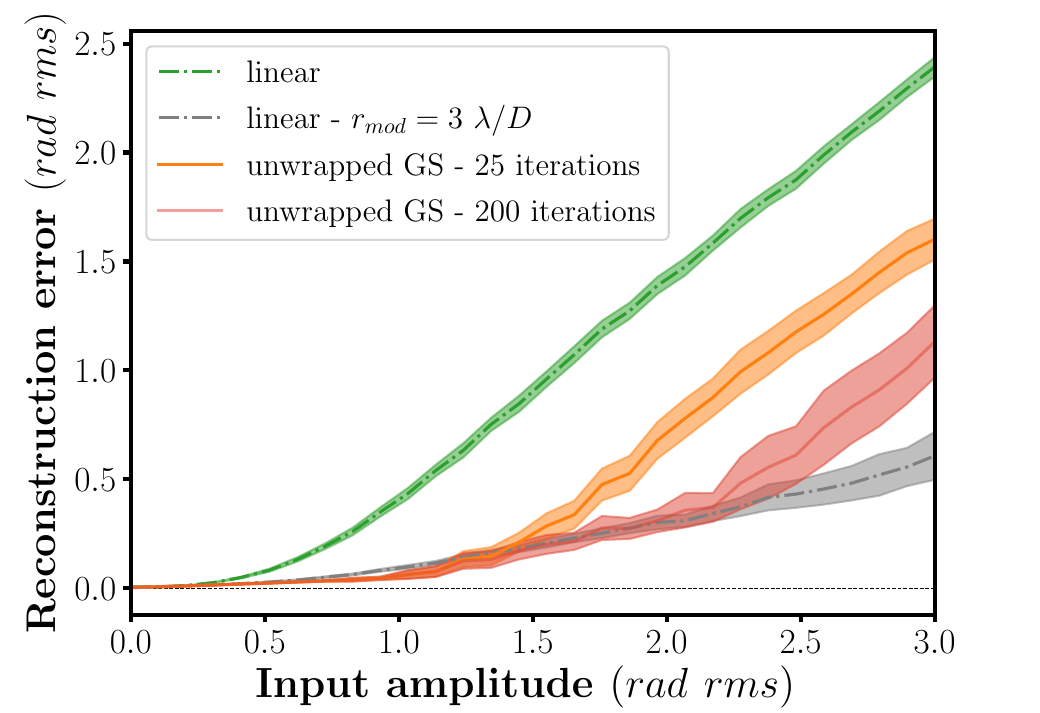}
\caption{Dynamic range plots produced by sending for each input amplitude 100 randomly sampled phase screen following a Von-Karman PSD and reconstructing them.}
\label{fig:linearity_scatter}
\end{figure}

Linearity curves provided in Figure \ref{fig:linearityCurves} and \ref{fig:linearityCurves_iterations} exhibit partial demonstrations of the gain in dynamic range, as they do not evaluate the potential non-linear cross talk between modes during reconstruction. Another approach to show the improved dynamics offered by the GS reconstructor is to generate dynamic range plots as proposed in \citep{PLWFS}. In this method, we introduce randomly sampled aberrations based on Von-Karman atmospheric Power-Spectral Density (PSD) into the PWFS system, varying the total wavefront error from 0 to 3 rad rms. For each wavefront error value, we inject and reconstruct 100 phase screens. This process is repeated for the nPWFS with the linear reconstructor, the GS algorithm with unwrapping (25 and 200 iterations), and the modulated PWFS ($r_{mod} = 3\ \lambda/D$). The mean values of the 100 reconstructions for each input amplitude across all configurations are plotted in Figure \ref{fig:linearity_scatter}, with the filled area representing the reconstructed error variance for each amplitude sample. It reveals that the GS algorithm outperforms the linear reconstructor for the nPWFS. When considering a satisfying reconstruction threshold as the point where reconstruction error reaches 10\% of the input rms, then the GS algorithm is extending the linearity range by a factor of approximately 3. Moreover, it achieves dynamic performance comparable to the $3\ \lambda/D$ modulated PWFS for input wavefront errors of up to 1.5 rad rms in the case of 25 GS iterations and up to 2.3 rad rms in the case of 200 GS iterations. For this test, the modulated PWFS shows best linearity as it is expected when analyzing linearity curves presented figure \ref{fig:linearityCurves}: modulated PWFS has the best dynamic range for low-order modes (which are affected by the modulation) and input turbulent phases used to produce curves Figure \ref{fig:linearity_scatter} have precisely higher amplitude for low order modes.\\

We have demonstrated that the GS algorithm combined with an unwrapping step can drastically improve the nPWFS dynamic range. Hence, it is possible to use this technique to reconstruct more accurately larger amplitude phases. Still, it is important to analyze how noise propagates through this reconstructor with respect to the linear framework, as one of the motivations to use the nPWFS is the better sensitivity with respect to noise.

\subsection{Noise propagation through reconstruction}

To investigate the way noise spreads across different reconstructors, the four configurations outlined in the preceding section will be used : a nPWFS with a linear reconstructor, a modulated PWFS, a nPWFS with a GS reconstructor without phase unwrapping, and a nPWFS with a GS reconstructor combined with phase unwrapping. For this study, only photon noise propagation will be analyzed, as it is a fundamental noise which cannot be mitigated by technological improvement. The  effect of noise propagation will simply be evaluated by introducing noise into the measurements and reconstructing the signals considering the different reconstructors. Once again, the idea is not to run an extensive simulation study on noise propagation through the GS algorithm in order to assess a large parameter space, but rather to highlight the basic sensitivity performance of this reconstructor with respect to the nPWFS and the modulated PWFS. To keep the analysis relatively simple, we study two configurations: noise impact on reconstruction of a flat wavefront (often referred to as reference intensities) and in the case of a turbulent phase screen with an amplitude of 0.75 radians rms (outside nPWFS linearity range but low enough to ensure that phase is not wrapped). To assess noise propagation relative to these configurations, the following procedure was held in simulation: for a given number of photons available for the measurement, the mean variance of each mode after reconstruction is estimated by averaging the variance of reconstructed modes over 200 noise realizations. This operation is repeated for different number of photons, setting the number of iteration in the GS algorithm to 25. This number of iteration was once again arbitrarily chosen as it was found that the number of iterations doesn't change the impact of noise propagation.

\begin{figure}[h!]
\centering
\includegraphics[width=0.5\textwidth]{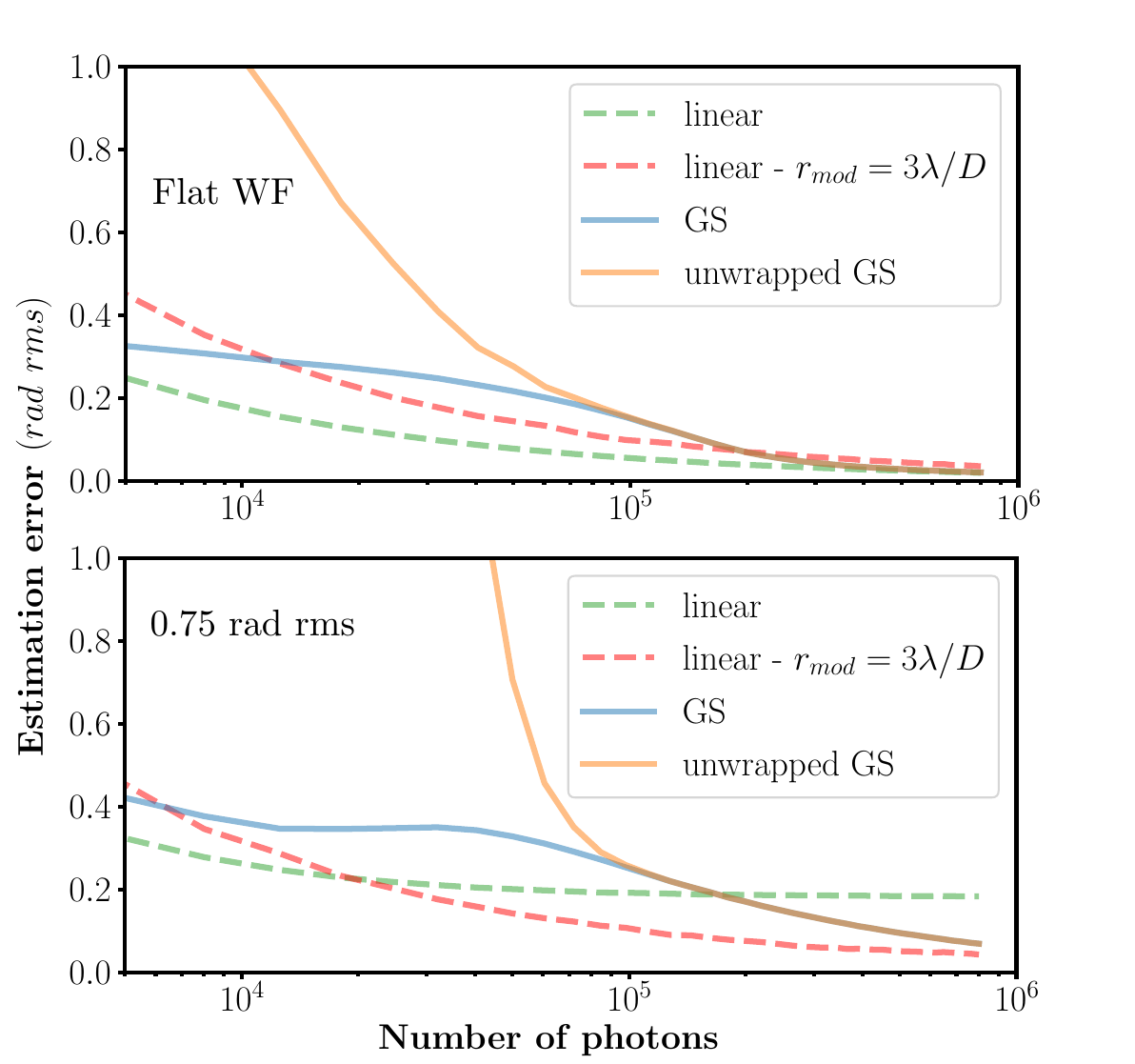}
\caption{Reconstruction errors due to photon noise. \textbf{Top} Case of a input flat wavefront. \textbf{Bottom} Case of an input turbulent wavefront of 0.75 radians rms.}
\label{fig:noise_propag}
\end{figure}

The phase reconstruction errors due to photon noise propagation in the case of a flat wavefront and a 0.75 radians rms turbulent screen with respect to incident flux are given figure \ref{fig:noise_propag}. The incident flux is expressed in total number of photons available for measurements assuming a prefect transmission and quantum efficiency of 1 for the detector (as a reminder: reconstruction is done over 500 Zernike modes).

For the cases of the linear reconstructor and a flat wavefront, the figure \ref{fig:noise_propag} (top) gives back a well-known behavior: the phase estimation error due to noise propagation through the reconstructor is lower for the nPWFS than for the modulated one. In the case of the GS algorithm, noise propagation behaves differently whether the unwrap step is done or not. 

Still in the context of noisy measurements for a flat wavefront: when not employing unwrapping, the GS algorithm propagates more error than the nPWFS and modulated PWFS for the high flux regime, and seems to perform slightly better than the modulated PWFS for the low flux regime. For even lower flux regimes (not shown here) GS would even appear to perform better than linear nPWFS reconstructor, but this behavior is misleading, in the sense that the GS algorithm will always provide wrapped measurements, preventing the amplitude of the reconstructed phase from diverging for extremely noisy measurements. Another way to explain this fact is to say that the GS comes with an intrinsic regularization that biases the results in this analysis. In the case where the wrapping step is added, the GS reconstructor builds more error in presence of noise compared to all the other studied configurations. 

Figure \ref{fig:noise_propag} provides the same analysis of noise propagation for the case of a turbulent phase screen (0.75 nm rms). We see the same trend, except that there is an offset for the linear nPWFS reconstructor which correspond to the reconstruction due to non-linearity error (not present for GS cases or the modulated PWFS).

\begin{figure}[h!]
\centering
\includegraphics[width=0.5\textwidth]{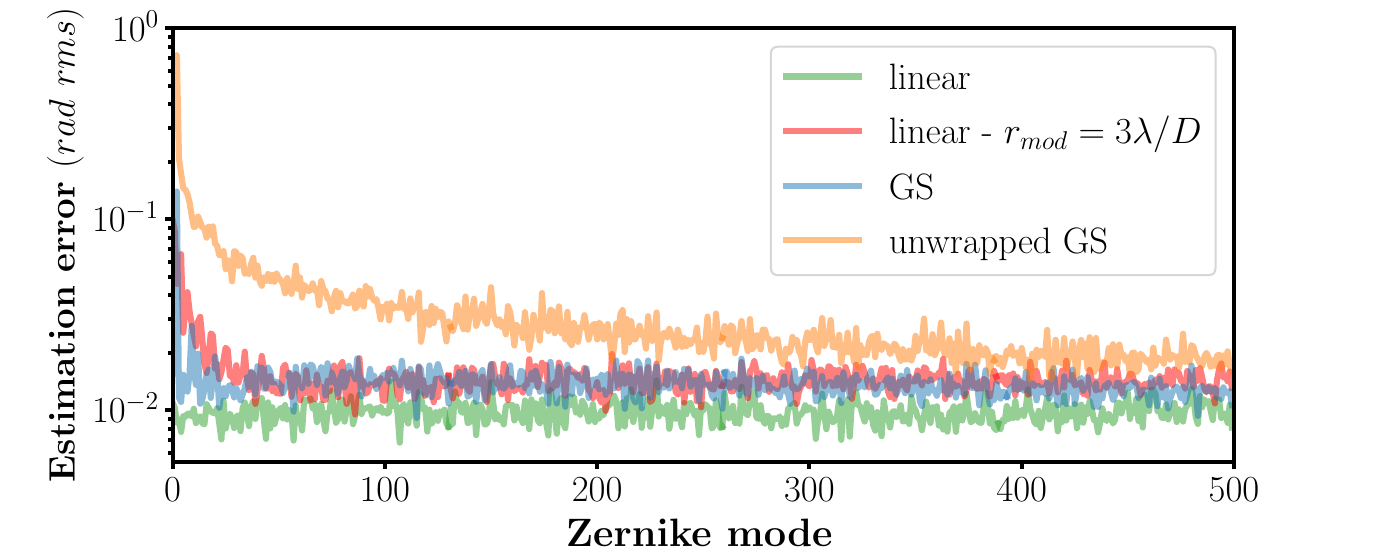}
\caption{Modal photon noise propagation for a case of 8000 photons available for the measurements.}
\label{fig:noise_propag_modes}
\end{figure}

To better understand noise propagation behavior, it is useful to check, for one flux regime, how the error propagates along the modes. Such an example is given figure \ref{fig:noise_propag_modes}, for the case of 400 photons available per mode. It is clear that the unwrapping step drives the noise to drastically propagate on the low-order spatial frequencies. 

To conclude this brief study on noise propagation, the GS reconstructor combined with an unwrapping algorithm performs badly in presence of noise as it propagates more noise than the modulated PWFS itself. Therefore, the sensitivity advantage of the nPWFS is lost while using this reconstructor. However, this implementation is for now in its most basic form. Noise propagation could be mitigated in the reconstructor through 2 aspects: on one hand by using a more noise-robust GS algorithm \citep{levin2020note}, and on the other hand by using an unwrapping algorithm with better behavior with respect to noise \citep{noiseUnwrap}. As mentioned in the introduction, using a nPWFS brings more advantages than only the gain in sensitivity with respect to noise, hence this current implementation of the reconstructor is still interesting for uses at high signal-to-noise ratio (SNR) (application examples are given in the conclusion). It is also important to notice that such a drawback for the GS algorithm was not raised in previous study proposing its implementation for reconstructing the curvature wavefront sensor signals \citep{guyon2010}.

\subsection{Broadband impact on performance}

It is crucial to take into account the potential influence of larger spectral bandpass measurements on the GS algorithm when reconstructing the EM field, as this technique assumes the coherence of the field. The nPWFS is achromatic in phase \citep{fauva2017}, meaning that it measures the phase of the incoming wavefront independently of the wavelength of the light (dispersion effects are neglected here). However, the amplitude of the wavefront does depend on the wavelength as aberrations present in the system and the turbulence usually introduce a fixed optical path difference (OPD) across all wavelengths. Hence, the measured signal will scale proportionally with the wavelength. This leads to 2 advantageous properties: for a flat wavefront, all the wavelengths will give the same measurements. In the right conditions, the closed loop can therefore converge towards a null-phase. Secondly, by choosing the central wavelength for the reconstruction, it allows the scaling of the phase to compensate between larger and smaller wavelength, giving back the monochromatic measurement in the case of the linear range (providing a flat spectrum). Overall, it is known that bandwidth has a limited impact on nPWFS measurements in general, and therefore it seems reasonable to expect the same for the GS reconstructor despite the fact that it assumes monochromatic measurements.

\begin{figure}[h!]
\centering
\includegraphics[width=0.4\textwidth]{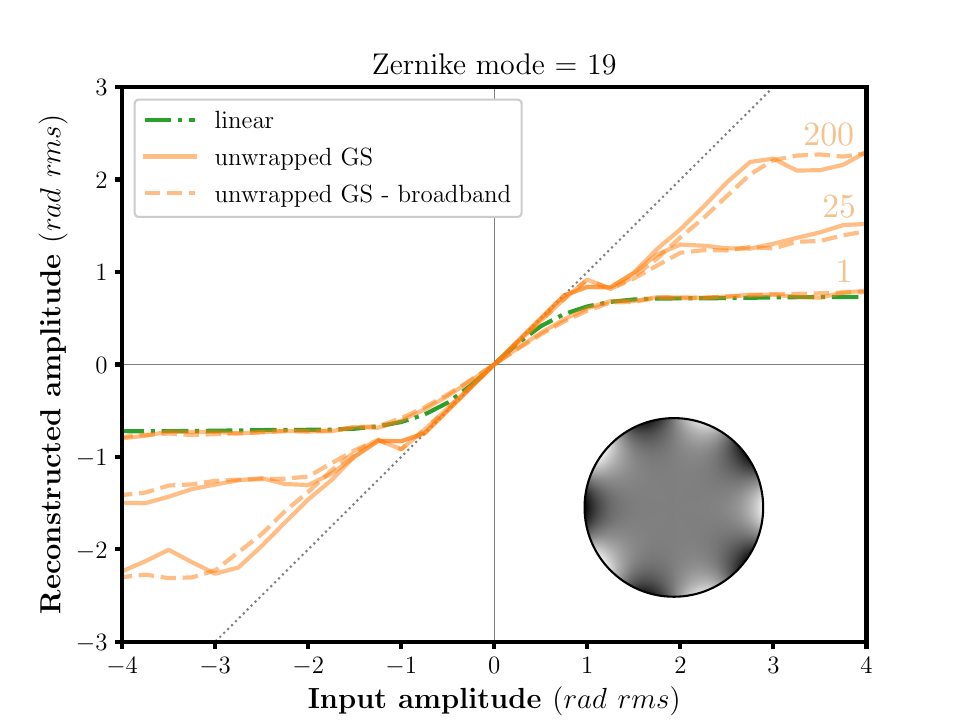}
\caption{Impact of broadband measurement on linearity curves for Zernike mode 19.}
\label{fig:Z19_broadband}
\end{figure}

A thorough investigation of the broadband impact on reconstruction would require an extensive study assessing various factors such as the amplitude of the measured phase, sampling used for the measurements, and the bandwidth. Due to the scope limitations of the paper, a brief example will be given to illustrate the impact of broadband measurement on the GS algorithm using the same configuration as in the previous section, but providing the algorithm measurements recorded by a polychromatic light having a flat spectrum ranging from $550\ nm$ to $800\ nm$ (bandwidth around 35\%, sampled with 50 points in our simulation). To run the GS algorithm, the nPWFS model is simulated as working at a monochromatic wavelength, set as the broadband central wavelength ($675\ nm$). Linearity curves for Zernike mode 19 in the case of different numbers of GS iterations are again displayed, but adding the one corresponding to the reconstruction of a broadband measurement (figure \ref{fig:Z19_broadband}). This figure demonstrates the limited impact of the broadband measurement on the reconstruction, despite the large bandwidth chosen.

\begin{figure}[h!]
\centering
\includegraphics[width=0.44\textwidth]{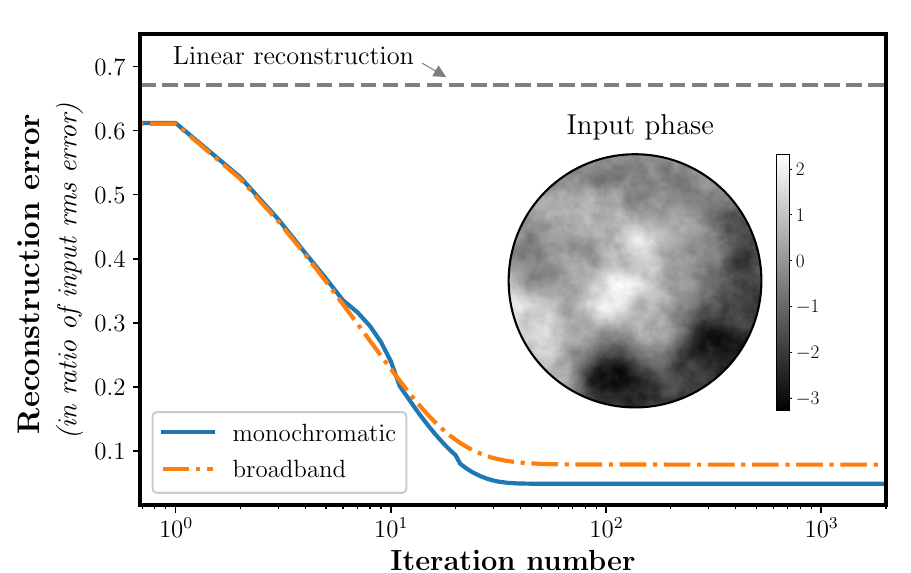}
\caption{Impact of broadband measurement on phase reconstruction. Input phase: turbulent screen for a case $D/r_{0}=4$ (colorbar in radians).}
\label{fig:convergence_broadband}
\end{figure}

To confirm the  GS algorithm robustness to broadband measurement, the reconstruction error as a function of iteration number for a turbulent screen in the case $D/r_{0}=4$ is plotted figure \ref{fig:convergence_broadband}. We see that the reconstruction error stabilizes to an reconstruction error slightly larger than the monochromatic case.

Overall, this short study points out that broadband measurements should not jeopardize the GS reconstruction scheme. We also mention that a polychromatic implementation of the GS algorithm could be imagined \cite{Fienup:99}, but it will comes at the expense of the computational time.

As the GS reconstruction scheme proposed in this paper is highly model-dependent, it is important to show that our technique is robust enough to model errors so it can be implemented on a real experimental setup. This will be achieved with the next section, in which we provide a experimental demonstration of our reconstructor.

\section{Experimental demonstration}

 The goal of this section is to present a laboratory demonstration of the GS algorithm achieved on the SEAL optical testbed at the University of California Santa Cruz. 
\subsection{Experimental setup}

\begin{figure}[h!]
\centering
\includegraphics[width=0.46\textwidth]{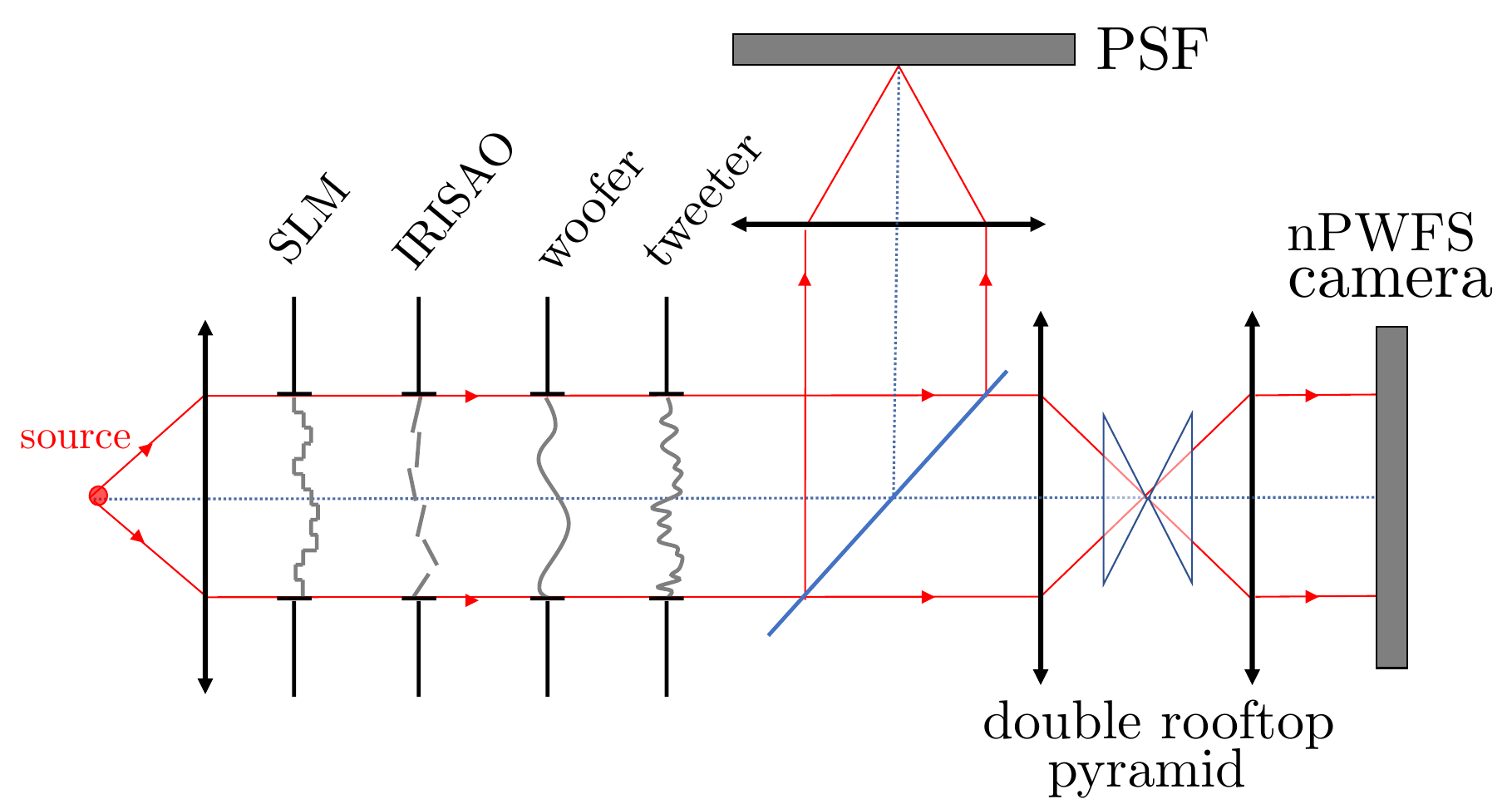}
\caption{A simplified layout of the SEAL testbed for the nPWFS subsystem.}
\label{fig:SEAL_layout}
\end{figure}

SEAL is an extreme adaptive optics testbed composed of several deformable mirrors (DM), wavefront sensors and coronagraphic branches \citep{SEAL}. A schematic layout of the SEAL testbed for the nPWFS subsystem only is presented in figure \ref{fig:SEAL_layout}. The SEAL components relevant for the experiment described here are the following ones:
\begin{itemize}
    \item[$\bullet$] Source at $\lambda=635\ nm$.
    \item[$\bullet$] Spatial Light Modulator (SLM): 1100 pixels across pupil diameter \citep{slmSEAL}.
    \item[$\bullet$] IRISAO Segmented deformable mirror (DM): ~6 segments across pupil diameter.
    \item[$\bullet$] Low-order ALPAO DM: 9 actuators across pupil diameter.
    \item[$\bullet$] High-order BMC DM: 24 actuators across pupil diameter.
    \item[$\bullet$] Focal plane camera at 2.3 Shannon sampling.
    \item[$\bullet$] Double rooftop nPWFS \citep{lozi2019visible} with 106 pixels across pupil diameter (same sampling as the simulations shown in this paper).
\end{itemize}

To build a reliable model of the SEAL nPWFS and run the GS algorithm, two quantities are required: an image of the pupil and the shape of the pyramid mask. To obtain an image of the pupil, large tip-tilt offsets are applied on the ALPAO DM in order to move the PSF away from the pyramid tip ($\sim20 \lambda/D$) and place it in each nPWFS quadrant successively. By doing so, four images of the pupil are obtained through each side of the pyramid mask. The SEAL pupil is then computed by re-centering and averaging these four images, with an estimated accuracy below one pixel. An image of the SEAL pupil measured through this method is given on the top-left of figure \ref{fig:GS_algo}, the segments gaps coming from the IRISAO DM are clearly visible. In order to compute the pyramid shape, the four pupil images are simply registered in order to produce the corresponding tip-tilt for each face of the pyramid mask. The phase to DMs (ALPAO and BMC) registration was done the following way: the central actuator of the DM is pushed and reconstructed and then pulled and reconstructed again. Difference of the images corresponds to the phase of this actuator. Then a waffle pattern is sent on the DM and reconstructed in order to register actuators positions. Finally the phase of the central actuator is fitted with a Gaussian function and duplicated at the other actuators positions. This calibration process requires only 4 images (2 for the central actuator and 2 for the waffle). In both cases of the ALPAO and BMC DM, the phase reconstructed is largely over-sampled (106 pixels across versus 23 actuators for the BMC). The nPWFS and associated GS algorithm is routinely used on the SEAL testbed with both DMs as the standard way to flatten the wavefront in order to correct for quasi-static aberrations, achieving a wavefront error of about 17 nm rms (as measured by the nPWFS).

\begin{figure}[h!]
\centering
\includegraphics[width=0.46\textwidth]{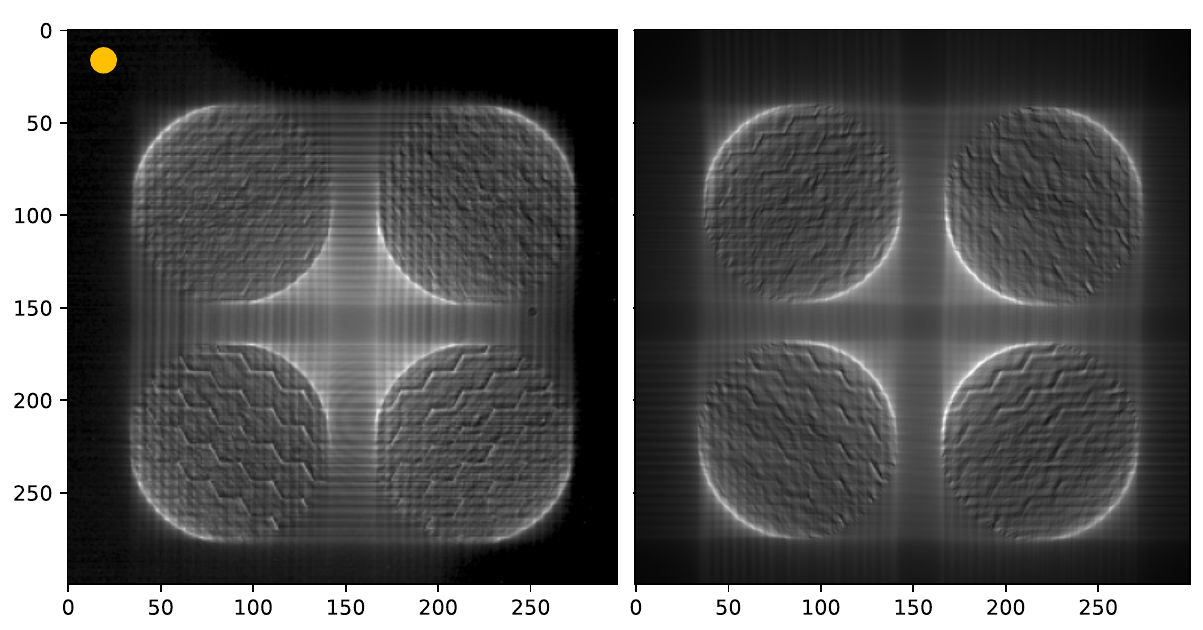}
\caption{\textbf{Left:} nPWFS image on SEAL after closed loop. nPWFS estimated wavefront error is 17 nm rms. \textbf{Right} Corresponding simulated image. The same scaling is used for both images.}
\label{fig:SEAL_img0}
\end{figure}

The nPWFS image after closed loop is given in Figure \ref{fig:SEAL_img0} (left). A high frequency pattern coming from the BMC DM is clearly visible (a well known effect, called the print-through or quilting effect). The corresponding residual phase is reconstructed from the reference image (best flat after closed loop on static aberrations) and propagated in the model. The obtained image is displayed Figure \ref{fig:SEAL_img0} (right), showing that the model exhibits a high fidelity with measurements. Among the small differences that can be spotted: print-through effect seems slightly underestimated in the simulated image (most likely because field-of view of simulated nPWFS is smaller than the real one, acting like a spatial low-pass filter) and a faint ring pattern can be seen between the two bottom pupil images of the true image (most likely coming from a dust particle on the glass pyramid). It is worth mentioning that all the results presented in the following subsections are obtained for measurements done at high SNR.

\subsection{Linearity curves}

As a first demonstration of the GS algorithm performance on the SEAL testbed, some of the linearity curves obtained in simulation in the previous section are reproduced. This study is done the following way: Zernike modes are sent with the SLM, in order to have a good knowledge value of the input phase. Following the same procedure as before, the linear reconstructor is calibrated with the Zernike modal basis, hence the reconstruction directly gives the value of each of the modes. For the GS algorithm, the phase is reconstructed for each pixel and then projected on the Zernike basis.

\begin{figure}[h!]
\centering
\includegraphics[width=0.4\textwidth]{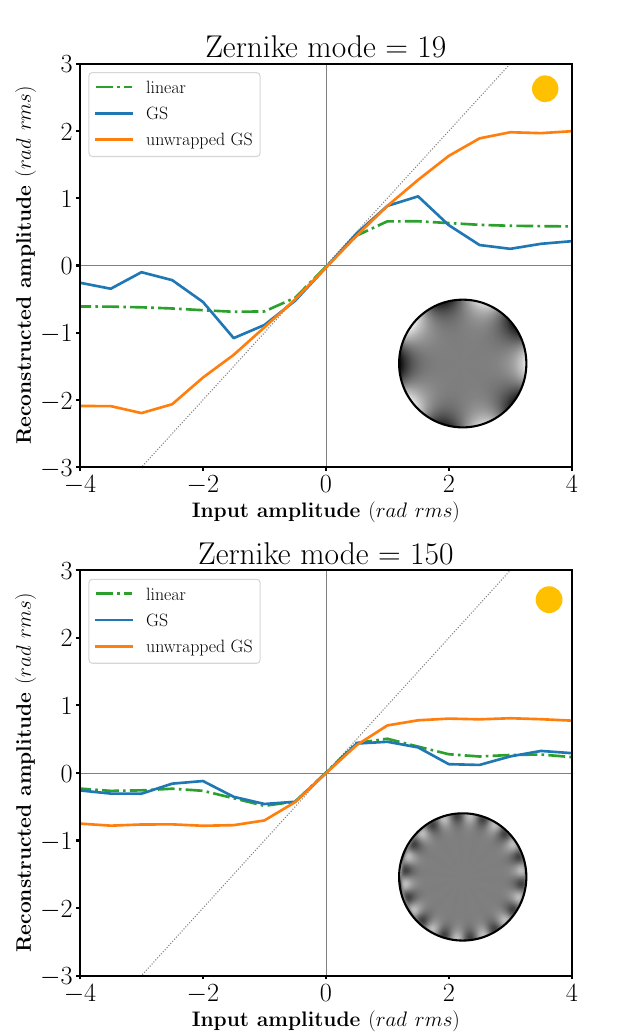}
\caption{\textbf{Top:} Linearity curve for $Z^{19}$ obtained on SEAL. \textbf{Bottom:} Linearity curve for $Z^{150}$ obtained on SEAL. GS algorithm is used with 200 iterations. Yellow dots are highlighting the fact these curves were obtained on the SEAL testbed.}
\label{fig:SEAL_linearity}
\end{figure}

Linearity curves obtained for the Zernike mode 19 and 150 are given in Figure \ref{fig:SEAL_linearity}, in the case of 200 iterations used for the GS algorithm. The conclusions from the simulations are confirmed in this experiment: the GS demonstrates a higher dynamic range than the linear reconstructor. However, for higher order modes, the GS algorithm without phase unwrapping seems to under-perform compared to simulation (Figure \ref{fig:linearityCurves}). This could be explained by model errors which are slightly affecting performance.

\subsection{Example of phase reconstruction and close loop} \label{section:seal}

\begin{figure}[h!]
\centering
\includegraphics[width=0.48\textwidth]{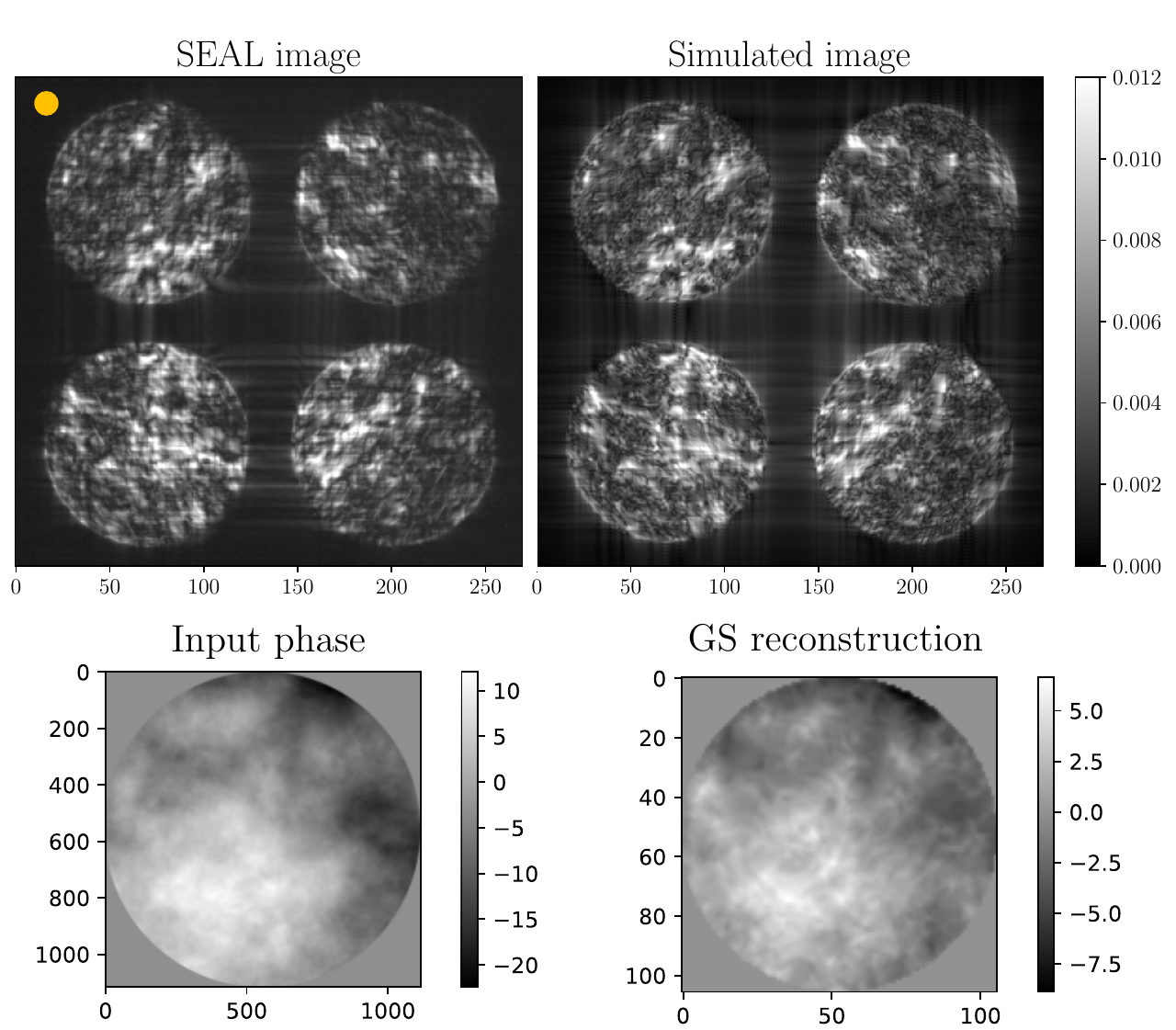}
\caption{\textbf{Top-left:} nPWFS image on SEAL testbed.\textbf{Top-right:} nPWFS simulated image (propagating reconstructed phase through nPWFS model). \textbf{Bottom-left:} Input phase on SLM. \textbf{Bottom-right:} Reconstructed unwrapped phase with 200 iterations for the GS algorithm.}
\label{fig:SEAL_turbu}
\end{figure}

As a demonstration of GS capabilities to reconstruct strong phase aberrations on the bench, a $D/r_{0} = 32$ phase screen is generated on the SLM (bottom-left of figure \ref{fig:SEAL_turbu}). The corresponding nPWFS image recorded is given in the top-left of the same figure, and the reconstructed unwrapped phase after 200 iterations is displayed on the bottom-right corner. Figure \ref{fig:SEAL_turbu} also shows a simulated image of the nPWFS signal, produced by simply propagating the reconstructed phase in the nPWFS model. Despite the fact the simulated image and the real image are almost identical, the phase is largely underestimated (more than a factor 2, this effect was also observed in the simulations presented in the previous section). This is explained by the fact that the nPWFS saturates: for an increasing input phase, the signal reaches a point where it almost does not change anymore. A simple example of such saturation is the case of a tip-tilt aberration: once the PSF is moved several tens of $\lambda/D$ in one quadrant, only one pupil image of the nPWFS is illuminated. Displacing the PSF even further away will almost have no impact on the measurements, as the illuminated pupil image is already concentrating all the flux. The saturation is an important effect that will limit the reconstruction range for any kind of linear but also non-linear reconstructor, as it implies that two different phases can lead to the same measurements. Hence, the saturation seems to be an intrinsic limitation in the nPWFS measurement, and inverting more accurately large amplitude aberrations would require extra knowledge on the phase to be measured. A potential solution to even further improve the dynamic range of GS reconstruction will be sketched in the next section.

\begin{figure}[h!]
\centering
\includegraphics[width=0.5\textwidth]{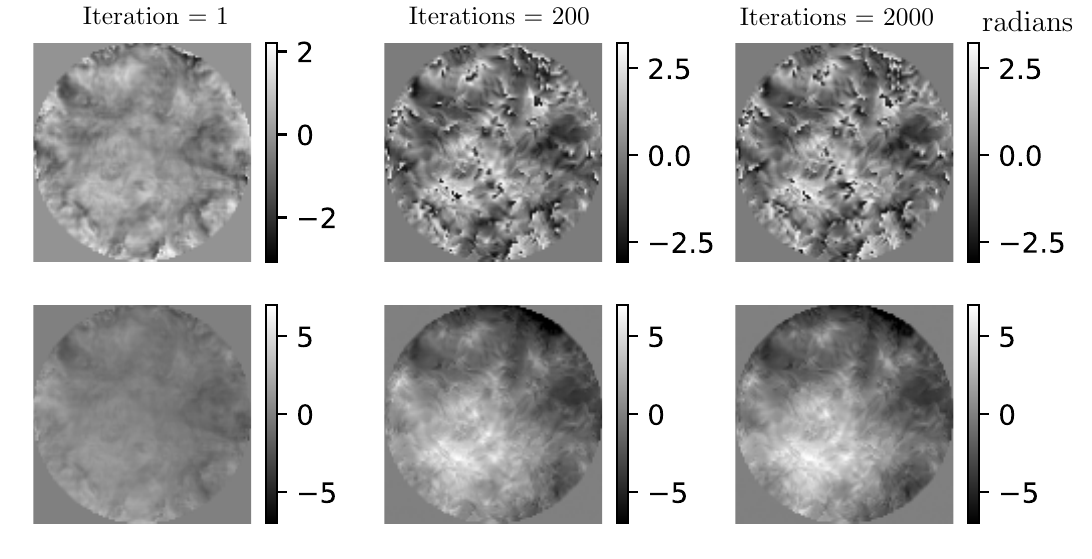}
\caption{\textbf{Top:} Wrapped phases reconstructed by the GS algorithm on SEAL. \textbf{Bottom:} Corresponding unwrapped phases.}
\label{fig:SEAL_iteration}
\end{figure}

Reconstructed phases at different iteration steps are displayed in figure \ref{fig:SEAL_iteration}. Once again the wrapped phases for 1, 200, and 2000 iterations are plotted on top row, and corresponding unwrapped phases are shown on the bottom row. As shown before, the phase estimation improves with iterations, but it seems that the estimation stops improving only after a few hundreds of iterations, instead of few thousands in simulations. Hence, the reconstructed phase for 200 and 2000 iterations are highly similar. Once again, this could be explained by the fact of differences between the nPWFS on the SEAL testbed and its model used for reconstruction.

\begin{figure}[h!]
\centering
\includegraphics[width=0.5\textwidth]{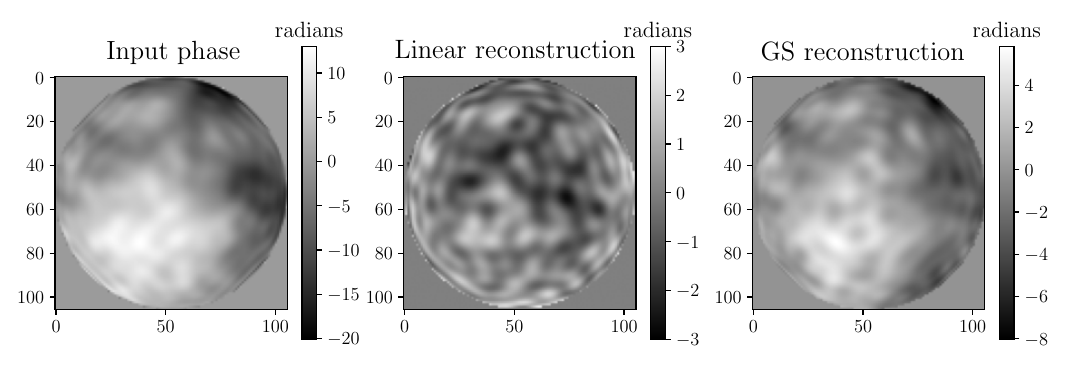}
\caption{\textbf{Left:} Input turbulent phase injected on the SLM in a case $D/r_{0} =32$, projected on the first 500 Zernike modes. \textbf{Middle:} Linear reconstruction. \textbf{Right:} GS reconstruction after 2000 iterations.}
\label{fig:SEAL_linearVSgs}
\end{figure}

It is also possible to compare the GS reconstruction with the linear reconstruction. To do so, a push-pull interaction matrix was measured on the bench by sending the first 500 Zernike modes on the SLM. The command matrix is then computed by taking the pseudo inverse of the interaction matrix and used to reconstruct the signal. The comparison between the linearly reconstructed phase and the GS reconstructed unwrapped phase projected on the first 500 Zernike modes are shown Figure \ref{fig:SEAL_linearVSgs}. It is clearly demonstrating that our GS reconstruction shows also a significant improvement compared to the linear reconstruction on the bench.

\begin{figure}[h!]
\centering
\includegraphics[width=0.49\textwidth]{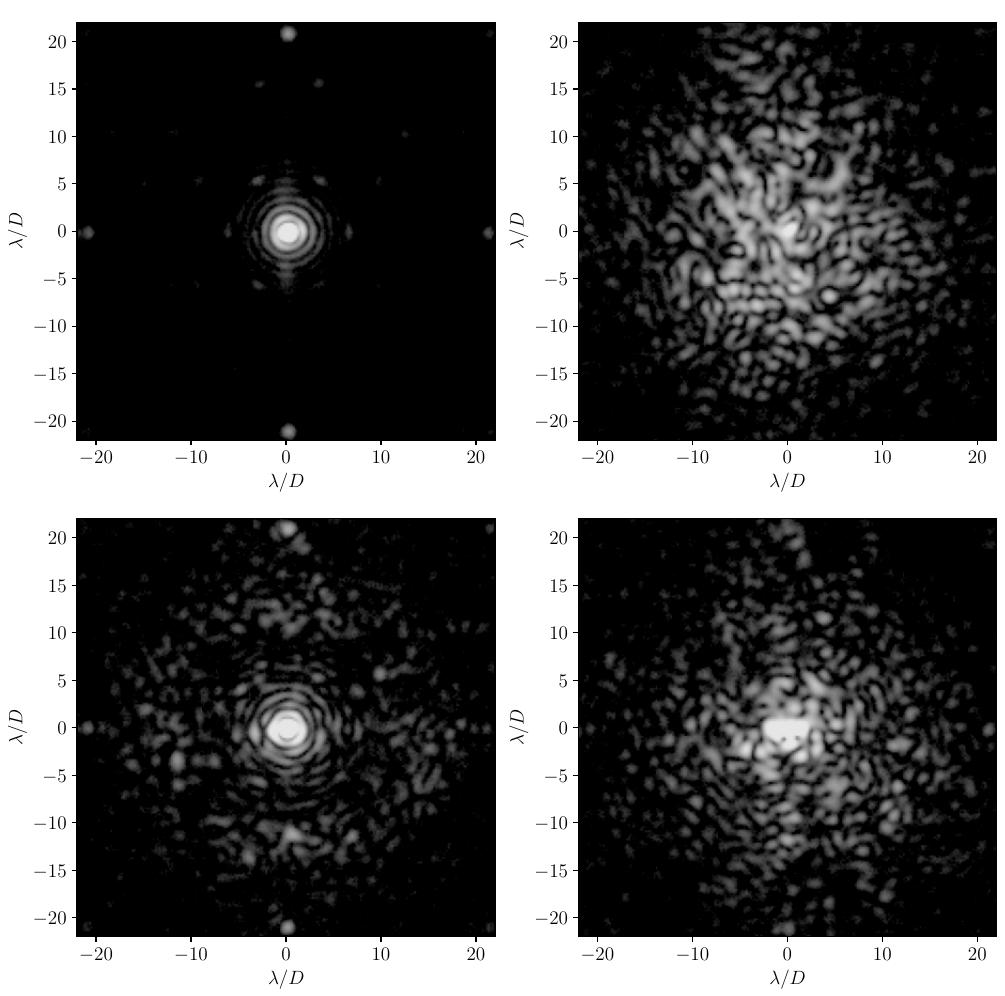}
\caption{Closing the loop with the BMC DM on a static atmospheric screen. \textbf{Top-left:} PSF corresponding to best flat on the SEAL testbed. \textbf{Top-right:} Uncorrected PSF for a input phase $D/r_{0}=32$ on SLM. \textbf{Bottom-left:} PSF after 12 closed loop steps using GS reconstructor. \textbf{Bottom-right:} PSF after 12 close loop steps using linear reconstructor.}
\label{fig:SEAL_psfs}
\end{figure}

To conclude the testbed demonstration, we present results of a closed loop run using the BMC DM (controlling all the modes) on the static turbulent screen presented in Figure \ref{fig:SEAL_turbu}. The linear reconstructor uses a push-pull zonal interaction matrix, and the GS reconstructor uses 25 iterations with phase unwrapping. For closed loop AO control, a simple integrator controller with a loop gain of 0.5 was used. The PSFs obtained after 12 closed-loop steps are presented in figure \ref{fig:SEAL_psfs}. In the figure, we also present the PSF corresponding to the best flat after closing the loop with the nPWFS on bench aberrations (with the BMC DM), and the uncorrected PSF corresponding to the atmospheric phase displayed on the SLM. The best-flat PSF exhibits a faint and vertical light pattern crossing its core. This comes from diffraction effects due to the SLM. The closed loop test clearly confirms the extended dynamic range of our reconstructor by showing that the closed loop with GS reconstructor out-performs the closed loop in the linear framework (which ends up diverging after a few tens of steps). It also shows that despite saturation effects, the AO closed loop scheme help to correct high amplitude phase with the nPWFS combined with GS algorithm. As the wavefront error drastically improve after bootstrap, we could imagine a closed loop scheme for which the number of GS iteration used for reconstruction are reduced once in the residual phases regime.

These experimental tests demonstrate the GS reconstructor performance on an optical bench, and show that it provides an improvement over the linear reconstructor.

\section{Beyond the non-modulated PWFS saturation}
\label{section:fGS}

\begin{figure}[h!]
\centering
	\includegraphics[width=1\columnwidth]{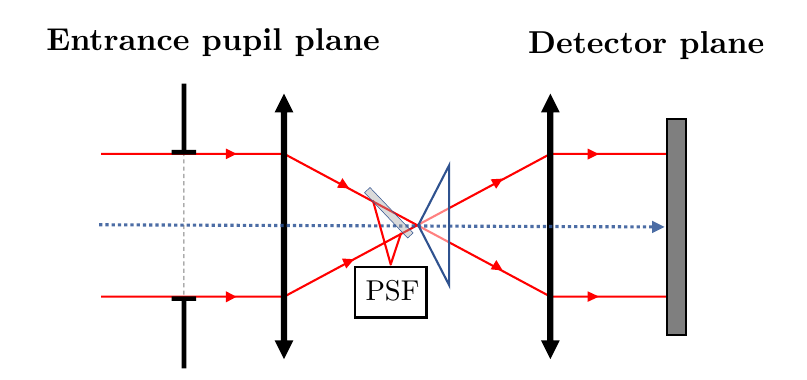}
    \caption{The focal plane assisted GS algorithm for the nPWFS.}
    \label{fig:GS_psf}
\end{figure}

As described in the previous section, the dynamic range of the GS algorithm seems to be limited by the saturation of nPWFS measurements. As shown, we can find different phases (in our case: the same shape, but different amplitudes) that give substantially the same measurements on the nPWFS. Therefore, high amplitude input phases bring a degeneracy in the measurements that seems to be unsolvable, no matter what non-linear reconstructor is considered. To push the dynamic range further, extra information on the phase is needed. To do so, the following strategy is proposed: using a second sensor that would be a focal plane camera located just before the pyramid tip, as shown in the schematic figure \ref{fig:GS_psf}.

A technique to use a focal plane image in order to push PWFS dynamics has already been proposed in \citep{2021A&A...649A..70C}. Here, the PSF image delivered by this camera provides the amplitude of the EM field in the focal plane, allowing this extra-information to be added in the GS algorithm procedure. In fact, the information of the EM amplitude recorded at the focal plane can be used to replace simulated focal plane EM field amplitude during back propagation (between step 1 and 2 figure \ref{fig:GS_algo}) and during forward propagation (between step 3 and 4 figure \ref{fig:GS_algo}). To do so, simulated amplitude of the EM field at the PWFS apex is replaced by the square root of the PSF measurement, while simulated EM phase is kept. We call this reconstruction the focal plane assisted GS (f-GS).\\

\begin{figure}[h!]
\centering
	\includegraphics[width=1\columnwidth]{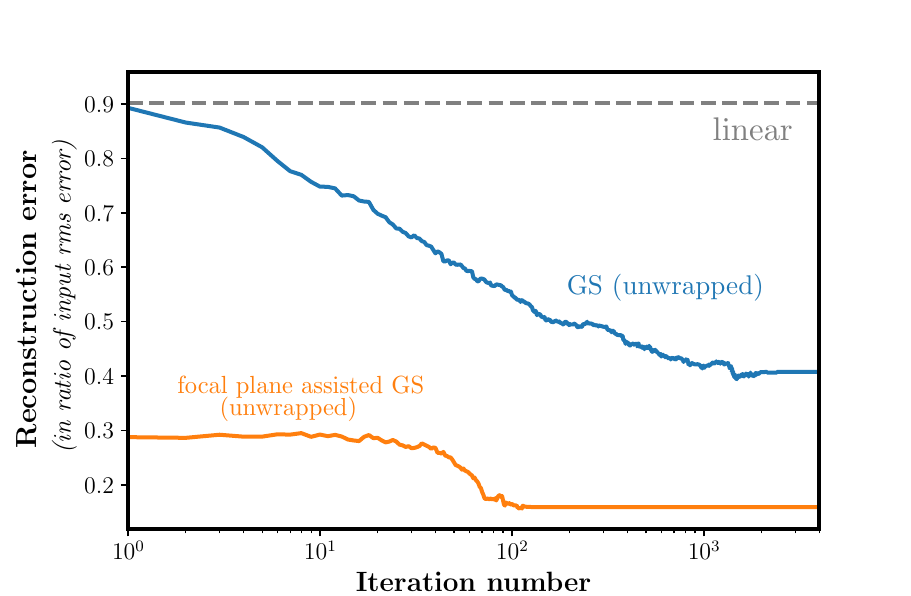}
    \caption{Simulated convergence plot for a input atmospheric phase in the case $D/r_{0}$=32. The focal plane assisted GS is outperforming the standard GS algorithm, both in convergence speed and final reconstruction accuracy.}
    \label{fig:GS_psf_convergence}
\end{figure}

\begin{figure}[h!]
\centering
	\includegraphics[width=1\columnwidth]{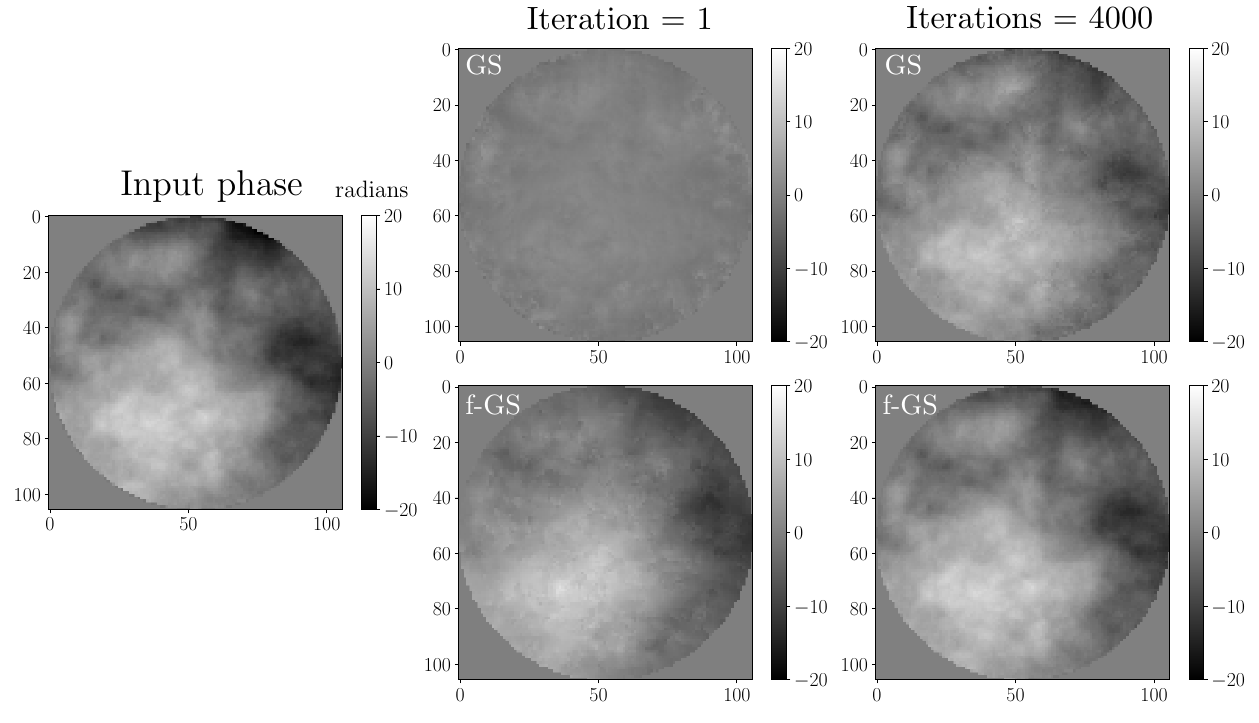}
    \caption{Reconstructed phases for the focal plane assisted GS compared to the standard algorithm, in the case of 1 iteration and 4000 iterations. All phases are displayed with the same scaling.}
    \label{fig:GS_psf_phases}
\end{figure}

The f-GS reconstructor was tested in simulation, using the same system as in section \ref{section:perfo}, using as input phase the atmospheric screen used in figure \ref{fig:SEAL_turbu} (case $D/r_{0}$=32). For this input phase, a convergence plot similar to figure \ref{fig:convergence} is presented in figure \ref{fig:GS_psf_convergence} for the case of the GS reconstructor and the f-GS reconstructor (both being unwrapped). The f-GS is outperforming the standard GS algorithm in two ways: first, the convergence speed is much faster and only one iteration is actually enough to provide a decent phase reconstruction. This represents a significant improvement as it could allow the GS approach to potentially be used in real time. Secondly, the overall reconstruction is also improved, showing that this strategy is indeed a solution to push the reconstruction further than the nPWFS saturation. The corresponding reconstructed unwrapped phases after 1 iteration and 4000 iterations for the GS and f-GS are given in Figure \ref{fig:GS_psf_phases}. It shows how after only one iteration the f-GS reconstructed phase is already really similar to the input phase, whereas the GS one is largely underestimated (f-GS providing a three times better reconstruction error on first iteration). Actually, even after 4000 iterations, the GS algorithm does not provide a better phase estimation than the f-GS after one iteration.

Hence, the f-GS could be a powerful tool to drastically increase the convergence speed of the algorithm and performance in the case of high amplitude aberrations. These advantages come with a more complicated practical implementation, requiring splitting the flux (an operation that could introduce non-common path aberrations) between two synchronized cameras. The details of such a implementation will be considered in future studies, with the main motivation of building a demonstration of the f-GS on the SEAL testbed. We precise that f-GS may not be needed in the case of a classic AO close loop scheme, as the nPWFS would work around residual phases and therefore far from the saturation regime after bootstrap is completed. Nevertheless, such a f-GS setup could open the nPWFS to a wider range of applications that would require to perform wavefront sensing on large phases.

\section{Conclusion}

This paper introduces a new way to invert the nPWFS measurements: it relies on the numerical model of the sensor and the use of the GS algorithm. We demonstrated that the GS based reconstructor along with an unwrapping algorithm can drastically improve the nPWFS dynamic range of the reconstruction compared to the linear framework, extending its linearity range by a factor of approximately 3. Moreover, it can achieve dynamic performance comparable to the $3\ \lambda/D$ modulated PWFS up to 2 rad rms. This technique was successfully demonstrated on the SEAL testbed at UCSC on which it was used to close the loop on high $D/r_{0}$ turbulent phase screens. This reconstructor, however, comes with two drawbacks: it has a high computational complexity which prevents it from being used for real time control purposes (in its current implementation) and noise propagation is worse compared to the linear reconstructor. However, noise propagation for only the most basic implementation of GS and phase unwrapping was studied in this paper. There might be a better suited GS algorithm combined with a noise-robust phase unwrapping algorithm that could improve the noise propagation. 
The GS reconstructor already has several areas which it can be useful in an AO system despite only being tested at high SNR at slow speeds. For example, calibration purposes (as  is done routinely on the SEAL testbed), segments/fragments phasing, highly sampled phase reconstruction, etc. It could also support a second stage nPWFS running in real time with a linear reconstructor through soft-real time reconstruction of the phase in order to compute optical gains, and could also be used to reconstruct telemetry data with higher fidelity.

The most promising path towards a real time implementation of this technique is using focal plane assisted GS, which we showed in simulation in section \ref{section:fGS}. This technique uses the fact that the GS algorithm is well suited for sensor-fusion and seems to significantly increase the convergence speed of the algorithm while allowing us to measure high amplitude phases with better accuracy. The next steps for this approach is to push the understanding of the f-GS and start to implement it on the SEAL testbed, using a PSF image as an extra-sensor.

Finally, the GS algorithm has been applied in this paper for nPWFS only, but we argue that it can be used for all Fourier-Filtering WFS and maybe beyond. In a more general way, this reconstructor belongs to a wider class of reconstructors: the ones using a numerical model of the sensor and iterative algorithms to increase the dynamic range.

\section*{Acknowledgments}
This work was performed under the auspices of the U.S. Department of Energy by Lawrence Livermore National Laboratory under Contract DE-AC52-07NA27344. The document number is LLNL-JRNL-850197.
%

\bibliographystyle{aa} 
\bibliography{biblio}

\end{document}